# Assessment of Transmission-level Fault Impacts on 3-phase and 1-phase Distribution IBR Operation

Qi Xiao, Jongha Woo, Lidong Song,
Bei Xu, David Lubkeman, Ning Lu
NC State University, Raleigh, NC, USA
{**qxiao3**, nlu2}@ncsu.edu

Abdul Shafae Mohammed,
Johan Enslin
Clemson University
North Charleston, SC, USA

Cara DeCoste Chacko, Kat Sico,
Steven G. Whisenant
Duke Energy
Charlotte, NC, USA

*Abstract*—**The widespread deployment of inverter-based resources (IBRs) renders distribution systems susceptible to transmission-level faults. This paper presents a comprehensive analysis of the impact of transmission-level faults on 3-phase and 1-phase distribution IBR operation. To evaluate distributed IBR tripping across various phases and locations on a distribution feeder, we conduct simulations of both symmetrical and unsymmetrical transmission faults at progressively greater electrical distances on a real-time transmission and distribution (T&D) co-simulation platform. The IBR power-to-load ratios (PLRs) at 50%, 100%, and 300% are considered to emulate low, medium, and high IBR conditions. Our results indicate that, while 1-phase and 2-phase faults typically trigger fewer IBR trips when compared to 3-phase faults, a significant power imbalance arises from the tripping of 1-phase IBRs on the affected phases. The imbalance can result in significant power quality problems and unintended equipment tripping. It may be necessary to design fault-ride-through mechanisms specifically tailored to 1-phase IBRs to help mitigate the power imbalances caused by unbalanced faults.**

## I. INTRODUCTION

In recent years, there has been a notable upsurge in the integration of distributed inverter-based resources (IBRs) within the electric power distribution system. A distribution feeder has the capacity to accommodate multiple MW-level, 3-phase solar farms, in addition to hundreds of kW-level, 1-phase rooftop photovoltaic (PV) systems. Additionally, there is a growing trend among both utilities and customers to adopt 1-phase and 3-phase battery energy storage systems (BESSs) for load shifting, PV output smoothing, voltage regulation, and to offer operational support for microgrid applications [1-3]. In the decades to come, the integration of distributed IBRs continues to be a crucial strategy to achieve 100% renewable energy by 2050.

When integrated into the main grid, IBRs are required to endure voltage and frequency disturbances through fault ride-through (FRT) functions, as specified by standards such as IEEE 1547-2018 [4]. However, although FRT helps IBRs withstand minor grid disturbances, it can lead to a significant number of IBRs disconnecting from the main grid during severe or prolonged disruptions. This was demonstrated in events like the wind and PV tripping events during the 2016 South Australia blackout [5]. This exacerbates the generation shortage and can potentially lead to cascading failures, especially during periods when the IBR power-to-load ratio (PLR) is high. Note that in this paper, PLR is defined as the combined power output of IBR divided by the total power consumption of loads, encompassing all battery charging loads.

Additionally, utility engineers have reported cases of IBR tripping during operational conditions, frequently triggered by short-lived voltage fluctuations following temporary interruptions or circuit-switching events. This significantly diminishes the reliability of power distribution grid operation. Moreover, fault currents injected by IBRs from the distribution grid into the transmission system can disrupt the known positive, negative, and zero-sequence current injection patterns, thus rendering the current protection settings ineffective for fault tripping.

Co-simulation [6-7] is often required for quantifying the impacts of integrating a large amount of distributed IBR on transmission and distribution (T&D) operations. In [7], the authors proposed a simplified transmission model (modeled in InterPSS) to simulate various transmission-level faults on IBR tripping in the distribution systems (modeled in OPENDSS). Nevertheless, the current co-simulation approach has a notable limitation: it inadequately represents the negative and zero sequence components contributed by IBRs in diverse operational conditions. This is particularly true when they operate in different grid-forming (GFM) and grid-following (GFL) control modes. The default IBR models lack the capability to represent various GFM and GFL control functions. Additionally, the use of phasor-domain IBR models proves inadequate in capturing responses in the Electromagnetic Transient (EMT) domain.

Therefore, in this paper, we investigate the effects of transmission-level faults on distributed IBR by utilizing an integrated T&D model on the OPAL-RT real-time simulation platform. Our primary contribution lies in enhancing modeling accuracy for negative- and zero-sequence current injections. This is achieved by incorporating distributed MW-level PV and BESS into the EMT domain. This allows for a detailed representation of customizable IBR control functions, including GFM, GFL, and FRT, as well as various grounding mechanisms. Furthermore, we systematically analyze the impacts of symmetrical and unsymmetrical transmission-level faults on distribution-level IBR tripping, considering factors such as phases, IBR locations, and IBR PLRs.

Our findings reveal that 1-phase and 2-phase faults, although initially causing fewer IBR trips compared to 3-phase

This research is supported by the Center for Advanced Power Engineering Research (CAPER) and the U.S. Department of Energy's Office of Energy Efficiency and Renewable Energy (EERE) under the Solar Energy Technologies Office Award Number DE-EE0008770.

faults, induce considerable power and voltage imbalances. These imbalances may subsequently lead to tripping events. Consequently, implementing a FRT mechanism tailored for 1-phase IBRs to address phase imbalances resulting from unbalanced faults may be deemed necessary.

## II. METHODOLOGY

In this section, we introduce the co-simulation testbed setup for modeling faults, IBRs, and FRT functions.

### A. Testbed Setup

As shown in Fig. 1, we run an integrated T&D network model on the OPAL-RT real-time simulation platform for fault analysis. The transmission model comprises a 3-phase, 1-machine, 6-bus configuration, where Bus 3 serves as the Point of Common Coupling (PCC). Connected to this configuration is a 3-phase, 123-bus distribution feeder characterized by a high penetration of IBRs.

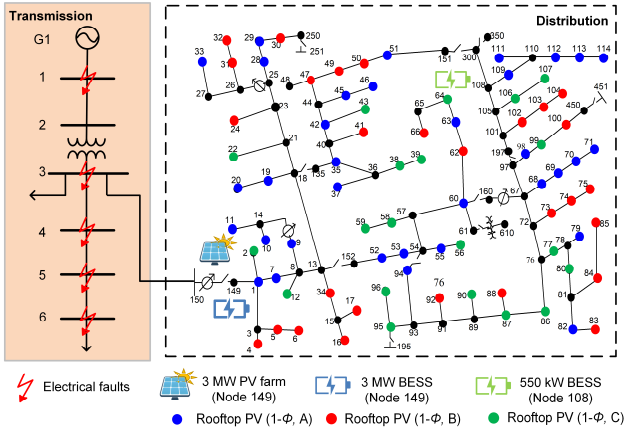

Fig. 1. Layout of the integrated T&D co-simulation testbed.

We model five fault locations: Generator Terminal Fault (GTF) at Bus 1, PCC bus fault at Bus 3 (PCC), and downstream faults at Buses 4, 5, and 6—denoted as SHORT, MEDIUM, and FAR, respectively. The three downstream fault cases correspond to short, medium, and long electrical distances from the PCC bus.

Please note that in this paper, our primary focus is to assess the impact of different types of transmission-level faults on a single high-IBR feeder, considering both upstream and downstream fault locations. Thus, we simplify the external transmission network to an equivalent generator with an internal impedance. In our follow-up journal paper, we plan to expand our analysis to include a complex transmission network with multiple high-IBR feeders connected to different locations.

### B. Distributed IBR Modeling

The distribution system comprises a 3-phase, 123-bus feeder model. The total feeder load varies between 1.3 MW and 3.5 MW, with the nodal loads being represented using ZIP load models. The 3-phase IBRs within this system include a 3-MW PV farm co-located with a 3-MW grid-following BESS, which is connected to node 149 at the head of the IEEE 123 bus feeder. Additionally, a standalone grid-following BESS with a capacity of 550 kVA, 1000 kWh is connected to node 108. All 3-phase IBRs are modeled in the EMT domain [8] using eMEGASIM. We place 86 1-phase rooftop PVs with a total power of 4.5 MW evenly throughout the feeder and across phases *a*, *b*, and *c*.

All 1-phase rooftop PV systems are modeled in the phasor domain using ePHASORSIM on the OPAL-RT platform, and their instantaneous output current is limited to twice the nominal value during faults. This configuration enables us to examine the effects of transmission-level faults on both 3-phase and 1-phase IBRs across various feeder locations and different phases. Furthermore, using the EMT model enables accurate representation of the IBR control functions, while incorporating grounding mechanisms allows us to depict the negative- and zero-sequence network components.

### C. Fault Ride-Through (FRT) Function Modeling

The IBR FRT functions are designed in compliance with IEEE 1547-2018 and PRC-024-2-Eastern Interconnection Standard [10]. In Fig. 2, the purple lines delineate the IBR FRT non-tripping zone. Within this zone, the IBR will maintain its operation when voltage and frequency fluctuate. If an IBR detects voltage or frequency outside the non-tripping zone, it will undergo a tripping mechanism and remain offline for a specified duration before attempting to reconnect to the grid.

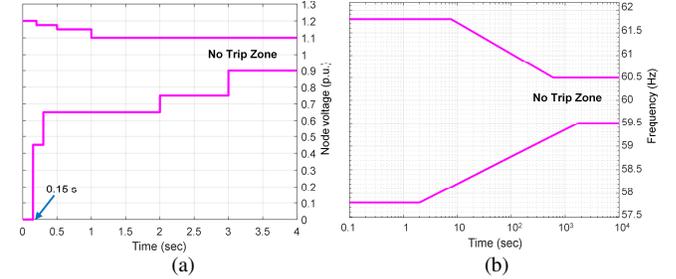

Fig. 2. IBR FRT functions. (a) Voltage FRT and (b) Frequency FRT.

### D. Transmission Faults Modeling

In this paper, constrained by page limits, we only report results for the following short-circuit fault types: single line-to-ground (SL2G), double line-to-ground (DL2G), line-to-line (L2L), and three-phase to ground faults (3L2G). The five fault locations have been described in Section II.A. The fault duration is uniformly set at 15 cycles because 9 cycles (i.e., 0.15 s) are selected as the smallest clearing time in FRT settings [10], as depicted in Fig. 2 (a).

The transmission system is represented in the sequence domain and the corresponding parameters are detailed in Table I. Note that $S_N$ is the nominal power, $V_{ll}$ is the line-to-line voltage; $Z_{short}$ is the short-circuit impedances; $l_{x\text{-}y}$ denotes the equivalent length from Bus *x* to Bus *y*. $Z_p$, $Z_n$, and $Z_0$ represent the positive-, negative-, and zero-sequence impedance of the line, respectively.

TABLE I. TRANSMISSION NETWORK PARAMETERS.

| Quantity | Value | Quantity | Value |
|---|---|---|---|
| $S_N$ | 10 MVA | $V_{ll}$ | 12.47 kV/4.16 kV |
| $Z_p, Z_n$ | 0.063 +*j*0.013 Ω/km | $Z_0$ | 0.016 +*j*0.041 Ω/km |
| $Z_{short}$ | 0.001 p.u. | $l_{1-3}$ | 9.2 km |
| $l_{3-4}$ | 1.5 km | $l_{3-5}$ | 4 km |
| $l_{3-6}$ | 7 km | | |

## III. SIMULATION RESULT ANALYSIS

In this section, we present an impact analysis of transmission faults at five locations along a distribution feeder that supports a substantial number of 3-phase and 1-phase distributed IBRs.

### A. Baseline Nodal Voltage Profiles

For a given IBR installed capacity, the IBR PLR typically varies between 10% to 300%. For instance, during a sunny hour with light loads, PLR may peak at 300%, while in an overcast hour with heavy loads, it can drop to 10%. In this paper, we set the total load at 1.3 MW and let both BESSs charge at their rated power (i.e., 300 kW and 100 kW). To simulate the feeder operating during a period of light load with passing clouds, we examine four IBR PLR scenarios: 0%, 50%, 100%, and 300%.

Please note that voltage regulation is not taken into account for the 0%, 50%, 100%, and 300% cases. This intentional omission is meant to assess the worst-case conditions without considering the impact of voltage regulation. However, to evaluate the influence of reactive power compensation on IBR tripping, we introduce a voltage regulation scenario (50%VR) for the 50% PLR case.

The pre-fault voltage profiles for the load nodes on phase $a$ without voltage regulation are illustrated in Fig. 3(a). As depicted in the figure, at 0% IBR PLR, the nodal voltage decreases with increasing electrical distance. However, as IBR PLR increases, the voltage drop along the feeder becomes less pronounced due to PV power injections. At 300% PLR, the IBR back feeds generated power into the transmission grid, causing certain distant nodes to register a voltage exceeding that of the substation.

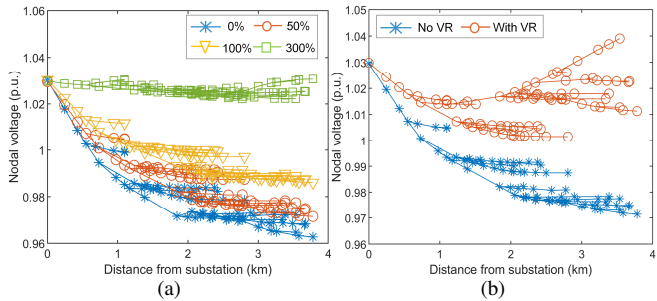

Fig. 3. Pre-fault nodal voltage profiles of phase $a$. (a) no voltage regulation for the four PLR cases (i.e., 0%, 50%, 100%, and 300%), and (b) with and without voltage regulation scenarios for the 50% PLR case.

As shown in Fig. 3(b), when shunt capacitors (located on nodes 83, 88, 90, and 92) and voltage regulators (nodes 150-149, 9-14, 25-26, and 160-67) are deployed, significant voltage increases are observed towards the end of the feeder. This is due to the combined impact of IBR real power and capacitor reactive power injections.

### B. Impact of Transmission Faults on Distributed IBRs

Table II summarizes the percentage of IBRs tripped offline for all four fault types. As mentioned in Section III.A, we modeled two cases for the 50% PLR scenario: one without voltage regulation (50%) and the other with voltage regulation (50%VR). For the 100% and 300% PLR scenarios, only cases without voltage regulation were modeled. In the table, '3-$\phi$' denotes the three-phase IBRs, including the 3-MW PV farm and the two BESSs; '$\phi$-A/B/C' represents the single-phase rooftop PVs on phases $a$, $b$, and $c$.

From the results, the following observations were made.

*1) **Impact of PLR:*** An increased PLR generally results in less IBR tripping. This is because the combined fault current injections from distributed IBRs to the transmission system can help raise the voltage levels along the distribution feeder, especially towards the feeder end, as shown in Figs. 4 and 5. The elevated voltage levels make it possible for the IBRs located at the end of the feeder to ride through the fault.

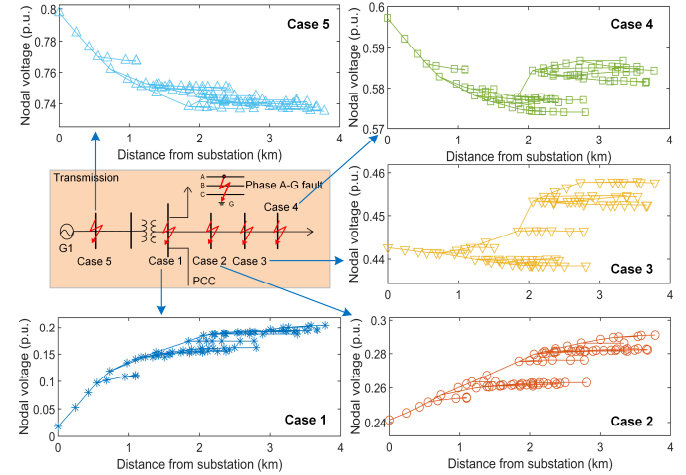

Fig. 4. Phase-$a$ nodal voltage snapshot during the SL2G fault (phase $a$ to ground fault) at 50% PLR and different fault locations.

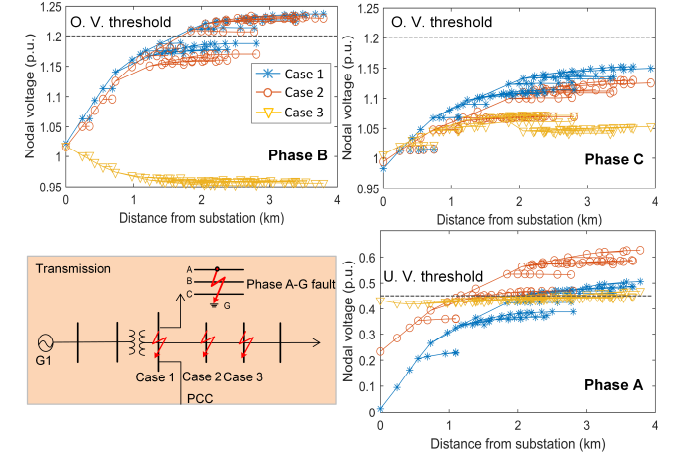

Fig. 5. Three-phase nodal voltage snapshot during the SL2G fault (phase $a$ to ground fault) at 300% PLR.

*2) **Impact of transmission fault locations:*** Among the five fault locations, the PCC fault results in the most pronounced voltage decline and leads to more IBR tripping, primarily because the distance to the fault location is the shortest. Meanwhile, downstream faults at the furthest location on Bus 6 result in no IBR tripping for all unsymmetrical faults and only partial tripping for three-phase-to-ground faults. This shows that positioning a high-IBR feeder closer to a stiff bus, where the voltage is closely regulated, can make it less susceptible to faults that happen in downstream circuits.

*3) **Impact of voltage regulation:*** In the two scenarios of 50% PLR, the injection of reactive current by voltage

TABLE II: Trip Percentage of Distributed IBRs Under Different Transmission Faults (IBR PLRs: 50%, 50%VR, 100%, 300%)

In each cell: the four numbers represent the results corresponding to the 50%, 50%VR, 100%, and 300% PV PLRs, respectively.

| Case | Fault Loc. | Single Line-to-Ground Fault (A-G Fault) | | | | Double Line-to-Ground Fault (A-B-G Fault) | | | | Line-to-Line Fault (A-B Fault) | | | | Three Phase-to-Ground Fault (A-B-C-G Fault) | | | |
|---|---|---|---|---|---|---|---|---|---|---|---|---|---|---|---|---|---|
| | | 3-φ (%) | φ-A (%) | φ-B (%) | φ-C (%) | 3-φ (%) | φ-A (%) | φ-B (%) | φ-C (%) | 3-φ (%) | φ-A (%) | φ-B (%) | φ-C (%) | 3-φ (%) | φ-A (%) | φ-B (%) | φ-C (%) |
| 1 | PCC | 100 | 100/100/100/62 | 0/0/0/68 | 0 | 100 | 100/100/100/57 | 100/100/100/55 | 0/0/0/41 | 0 | 0 | 0 | 0 | 100 | 100/100/100/57 | 100/100/100/50 | 100/100/100/63 |
| 2 | Down stream SHORT | 100 | 100/100/100/57 | 0/0/0/59 | 0 | 100 | 100/100/57/54 | 100/100/100/41 | 0/0/0/48 | 0 | 0 | 0 | 0 | 100 | 100/100/100/57 | 100/100/100/41 | 100/100/100/52 |
| 3 | Down stream MEDIUM | 0 | 0 | 0 | 0 | 100 | 97/57/14/0 | 100 | 0 | 0 | 0 | 0 | 0 | 100 | 100 | 100/100/100/41 | 100 |
| 4 | Down stream FAR | 0 | 0 | 0 | 0 | 0 | 0 | 0 | 0 | 0 | 0 | 0 | 0 | 33/33/0/0 | 95/35/3/0 | 59/47/0/0 | 89/0/0/0 |
| 5 | GTF | 0 | 0 | 0 | 0 | 100 | 0 | 100/100/100/82 | 0 | 100 | 0 | 100/100/100/59 | 0 | 100 | 100 | 100 | 100 |

regulation devices is observed to elevate nodal voltages, enabling some downstream single-phase IBRs to ride through the fault. This shows the effectiveness of voltage regulation in reducing IBR tripping.

*4) 3-phase IBRs versus 1-phase IBRs:* In general, 3-phase IBRs are more susceptible to transmission-level faults than 1-phase IBRs. This is because large 3-phase IBRs are typically placed at the feeder head, where the fault voltage is the lowest. Additionally, 3-phase IBRs can also be tripped as a result of voltage imbalance induced by unsymmetrical faults.

*5) 3L2G faults:* As expected, 3L2G faults consistently represent the most critical scenarios, leading to the highest number of IBR tripping incidents in nearly all cases. Notably, it is observed that under high IBR PLRs, certain feeder-end IBRs may withstand the faults due to the elevated voltage resulting from substantial IBR fault injections.

*6) SL2G faults:* In Fig. 4, during the phase $a$ to ground fault at the PCC bus (Case 1) with a 50% IBR PLR, phase $a$ voltage rapidly drops close to 0 p.u., leading to the tripping of all 3-phase and phase $a$ IBRs. Note that phase $b$ and phase $c$ rooftop PVs remain unaffected. At the 300% PLR for the same fault case, the phase $a$ IBR situated at the end of the feeder remains unaffected, as shown in Fig. 5. This is because the fault current injections from IBRs elevates the voltage towards the end of the feeder, making it surpass the under-voltage threshold. However, the elevated voltage on phase $a$ can also result in over-voltage on phase $b$ due to the increased unbalanced fault current, subsequently unintentionally tripping IBRs on those unaffected non-fault phases.

*7) DL2G faults:* We simulate DL2G faults by introducing ground faults on phases $a$ and $b$. Fig. 6 depicts the three-phase nodal voltage snapshot during the DL2G fault of Case 1 at 50% and 300% PLRs. As expected, 3-phase IBRs and IBRs on phases $a$ and $b$ are affected at 50% PLR. Similar to the SL2G fault, some IBRs on phases $a$ and $b$ at the end of the feeder survive due to the boosted feeder-end voltage at 300% PLR. Additionally, overvoltage on phase $c$ is observed due to the unsymmetrical fault on phases $a$ and $b$, resulting in the partial tripping of phase $c$ IBRs. However, in Case 5, the transformer connection between Buses 2 and 3 is a $\Delta$-$Yg$, in which the low-voltage winding ($Yg$) lags high-voltage winding ($\Delta$) by 30 degrees. Consequently, phase $b$ experiences a substantial voltage drop, even though both phases $a$ and $b$ are short-circuited at Bus 1. This results in the trips of all three-phase and phase-$b$ IBRs.

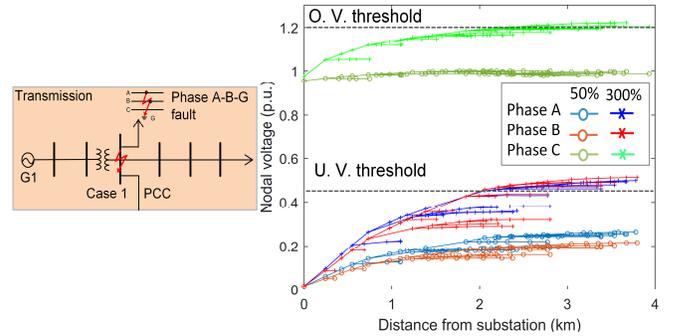

Fig. 6. Three-phase nodal voltage snapshot for the DL2G fault (phases $a$ and $b$ to ground fault) of Case 1 at 50% and 300% PLRs.

*8) L2L faults:* Line-to-line faults cause the least voltage drops along the transmission line. They exhibit minimal influence on distributed IBRs, resulting in no tripping for cases 1- 4. As depicted in Fig. 7, similar to DL2G in Case 5, a fault at the generator terminal induces a substantial voltage drop in only phase $b$ on the distribution feeder due to the shielding effect of the $\Delta$-$Yg$ transformer between Buses 2 and 3. Meanwhile, the voltages on phases $a$ and $c$ remain above the ride-through threshold. Consequently, only the IBRs on phase $b$ and the three-phase IBRs will trip. When comparing the two penetration scenarios (i.e., 50% vs. 300%), there are fewer IBR trippings at 300% PLR due to the elevated voltage caused by the increased contribution of IBR fault currents.

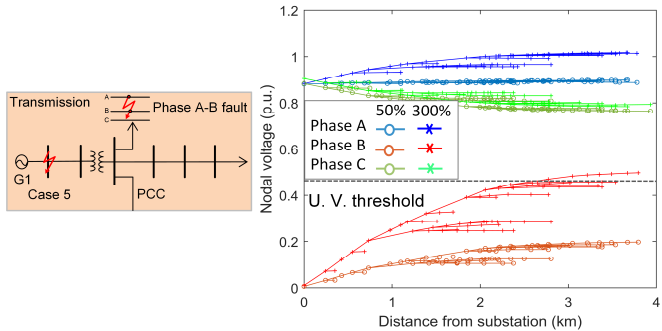

Fig. 7. Three-phase nodal voltage snapshot for the L2L fault (phases *a* to *b* fault) of Case 5 at 50% and 300% PLRs

*9) Voltage Unbalance:* All tripped-off IBRs must remain in the off state for a duration specified by the IEEE 1547-2018 standard. As a result, after an unsymmetrical fault, a large number of single-phase IBRs are tripped offline and remain off for a period, causing a substantial power imbalance among various phases. This power imbalance can subsequently lead to significant voltage imbalance.

In a distribution system, the severity of voltage imbalance is measured by the voltage unbalance factor (VUF), with an acceptable range set within 0.03 [11]. As depicted in Fig. 8, 1-phase (phase *a*) and 2-phase (phases *a*, *b*) line-to-ground faults represent the worst-case scenarios. The severity of voltage imbalance increases with higher IBR penetration. For example, the maximum UVF value exceeds 0.06 at 300% PLR. Note that the activation of voltage regulation devices (Fig. 8(a) vs. Fig. 8(b)) only marginally reduces the VUF because the location and quantity of reactive power injections are not optimized for per-phase voltage regulation.

Highly imbalanced phase voltage raises power quality concerns and may lead to inadvertent load and IBRs tripping. This result demonstrates the need for considering the impact of unsymmetrical faults on single-phase IBRs in the future design of FRT standards.

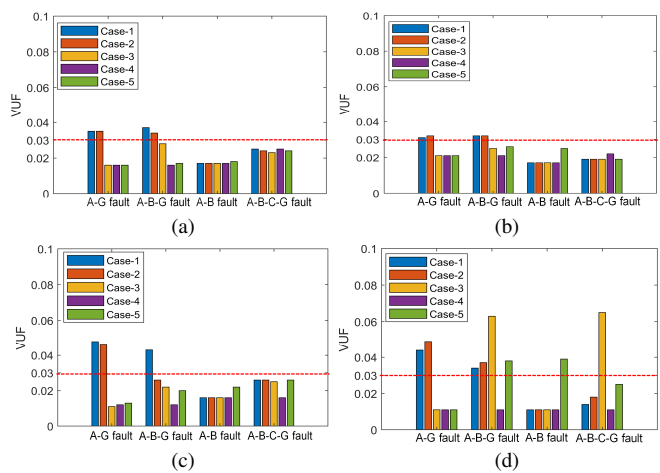

Fig. 8. Maximum VUF of distribution nodes at (a) 50%, (b) 50%VR, (c) 100%, and (d) 300% PLRs after different faults.

## IV. CONCLUSIONS

In this study, we conducted a comprehensive analysis of the influence of different types of transmission faults on a distribution feeder with high IBR penetration. We modeled different transmission fault locations and types, distributed IBR locations and types, and various IBR PLRs. The integrated T&D model is co-simulated on the OPAL-RT real-time simulation platform, where 3-phase IBRs are modeled in the EMT domain and 1-phase IBRs are modeled in the phasor domain.

Our research findings suggest that in cases where the ratio of IBR power to the load is high, the voltage drops along the distribution feeder are notably impacted by the magnitude of the backfeeding current. This results in non-uniform IBR tripping on the distribution feeder, with IBRs located at the end of the feeder being more likely to remain operational during faults. Moreover, during unsymmetrical faults, the injected currents from IBRs can result in overvoltage in non-faulty phases, potentially causing the tripping of IBRs in those non-faulty phases. Lastly, both symmetrical and unsymmetrical faults can cause the unsymmetrical tripping of single-phase IBRs, leading to significant node voltage imbalances in high IBR penetration scenarios, even when the PCC voltage remains balanced.

Our future work will be focused on assessing how transmission faults impact multiple high-IBR feeders within a complex transmission network that includes transmission-level IBRs.